\documentclass{article}
\usepackage{preprint}

\usepackage{microtype}
\usepackage{graphicx}
\usepackage{subcaption}
\usepackage{booktabs} 

\usepackage{amsmath,amssymb}

\usepackage[square]{natbib}
\usepackage[utf8]{inputenc}	
\usepackage[T1]{fontenc}	
\usepackage{xcolor}		
\usepackage[colorlinks = true,
            linkcolor = purple,
            urlcolor  = blue,
            citecolor = cyan,
            anchorcolor = black]{hyperref}	
\usepackage{booktabs} 		
\usepackage{nicefrac}		
\usepackage{microtype}		
\usepackage{lineno}		
\usepackage{float}			

\usepackage{lipsum}		

\usepackage{newfloat}
\DeclareFloatingEnvironment[name={Supplementary Figure}]{suppfigure}
\usepackage{sidecap}
\sidecaptionvpos{figure}{c}

\usepackage{titlesec}
\titlespacing\section{0pt}{12pt plus 3pt minus 3pt}{1pt plus 1pt minus 1pt}
\titlespacing\subsection{0pt}{10pt plus 3pt minus 3pt}{1pt plus 1pt minus 1pt}
\titlespacing\subsubsection{0pt}{8pt plus 3pt minus 3pt}{1pt plus 1pt minus 1pt}

\usepackage{tikz,xcolor,hyperref}

\definecolor{lime}{HTML}{A6CE39}

\def\TD{\mathbf D}
\def\TM{\mathbf M}
\def\bh{\mathbf h}

\title{Inverting airborne electromagnetic data with machine learning}

\usepackage{authblk}

\author[1,2\thanks{\tt{Mike.mcmillan85@gmail.com}}]{Michael S. McMillan}
\author[2\thanks{\tt{bas@compgeoinc.com}}]{Bas Peters}
\author[2\thanks{\tt{ophir@compgeoinc.com}}]{Ophir Greif}
\author[2\thanks{\tt{paulina@compgeoinc.com}}]{Paulina Wozniakowska}
\author[3]{Eldad Haber}

\affil[1]{Invert Geophysics, Svolvaer, Norway}
\affil[2]{Computational Geosciences Inc. Vancouver, BC, Canada}
\affil[3]{University of British Columbia, Vancouver, BC, Canada}

\begin{document}

\maketitle

\begin{abstract}
This study focuses on inverting time-domain airborne electromagnetic data in 2D by training a neural-network to understand the relationship between data and conductivity, thereby removing the need for expensive forward modeling during the inversion process. Instead the forward modeling is completed in the training stage, where training models are built before calculating 3D forward modeling training data. The method relies on training data being similar to the field dataset of choice, therefore, the field data was first inverted in 1D to get an idea of the expected conductivity distribution. With this information, $ 10,000 $ training models were built with similar conductivity ranges, and the research shows that this provided enough information for the network to produce realistic 2D inversion models over an aquifer-bearing region in California. Once the training was completed, the actual inversion time took only a matter of seconds on a generic laptop, which means that if future data was collected in this region it could be inverted in near real-time. Better results are expected by increasing the number of training models and eventually the goal is to extend the method to 3D inversion.
\end{abstract}

\clearpage

\section{Introduction}
Airborne electromagnetic (AEM) data has traditionally been inverted in 1D \citep{Farquharson1996}, 2D \citep{Wilson2006} and most recently 3D \citep{Haber2007} in order to best interpret the returning responses emanating from the earth during an AEM survey. These inversion methods along with plate modelling and parametric inversion approaches \citep{McMillan2015} have been at the backbone of AEM interpretation for many years. The application of 3D physics-based inversion is costly however. The costs arise from repeatedly solving partial differential equations (PDE) at large scale in 3D as well as fine-tuning the required meshes, uncertainty estimates and regularization parameters. Recently, machine learning algorithms have started to tackle the AEM inversion probem in 1D \citep{Wu2022}, where expensive forward modeling during the inversion is avoided, but how well neural networks can generalize to 2D and 3D inversions is still a largely under-explored topic. Here we propose a novel approach to map AEM data to conductivity models along 2D flight lines using a deep neural network. We first demonstrate how to create training data, and then show how to train and validate the network, before ultimately testing the network predictions on a field dataset. 

\section{Method}

Our primary goal is to train a neural network that inverts AEM data  without solving PDEs. Instead, all PDE solves will be off-line in the creation of the training dataset using 3D forward modelling. We train the network using the forward modelled data and, once trained, the network can convert time-domain AEM field data into 2D conductivity models along flight lines in a matter of seconds.

The network operates image-to-image, that is, the input data is of size $\TD \in \mathbb{R}^{n_t \times n_s}$, where $n_t$ indicates the number of time-gates and $n_s$ indicates the number of stations. The output needs to be a conductivity model, an image, of size $\TM \in \mathbb{R}^{n_z \times n_s}$, where $n_z$ is the number of pixels in the depth direction, and $n_s$ is again the number of stations. AEM data is heavily influenced by the varying nature of flight heights, of size $\bh \in \mathbb{R}^{n_s}$, therefore, these flight heights are also included in the network. This is done via a learned embedding that converts the flight heights into an image. We emphasize that while the network operates image-to-image, it is an inversion equivalent because it learns how to map 3D forward modeled data into a 2D conductivity model.

The challenge of a PDE-free inversion via a network is twofold. The first problem is to generate training data and models that are sufficiently close to the field data such that the network will generalize. Secondly, the different dimensions of AEM data, flight heights and conductivity models prohibit the use of plain UNets \citep{Ronneberger2015}, ResNets \citep{He2016} or VNets \citep{McMillan2021}. We address these challenges in the sections below where the network is described as well as the training procedure.

\section{Network Architecture}

The network architecture is fully convolutional, and we propose a three branch network that takes care of merging flight heights with AEM data, mapping time-domain data into depth-domain feature images, and translating the features into a conductivity model.

Let $E (\bh,\theta_e)$ represent the embedding of the flight heights, let $F ((\TD, E (\bh,\theta_e)), \theta_f)$ represent the network that converts the data from time to depth, let $G(F ((\TD, E (\bh,\theta_e)), \theta_g))$ represent the translation from intermediate features to the final conductivity model. In the above definitions, $\theta_e, \theta_f, \theta_g$ denote the network weights (convolutional kernels) and biases. 

\begin{figure}[!htb]
	\centering
	\includegraphics[width=0.92\textwidth]{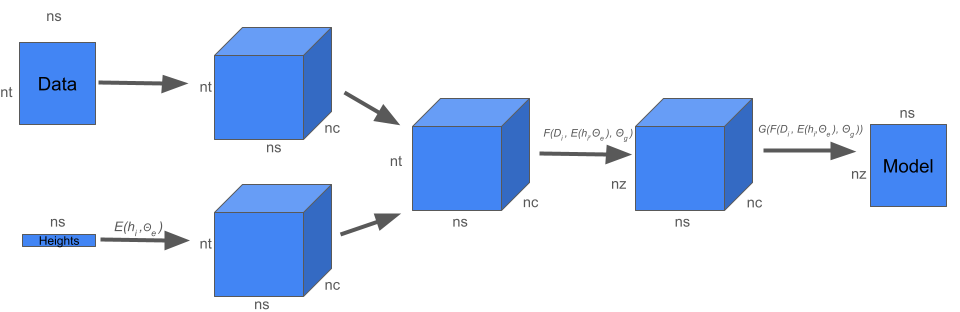}
	\caption{Simplified depiction of the neural network architecture, nt = number of time-channels, ns = number of stations, nc = number of channels in the network, nz = number of depth values.}
	\label{fig:network}
\end{figure}

Our approach trains all networks jointly, by minimizing the mean squared error with a loss function:
\begin{equation}
    L(\TD,\TM,\bh,\theta_e,\theta_f,\theta_g) =  \frac{1}{n_{ex}} \sum_{i=1}^{n_{ex}}\| G(F ((\TD_i, E (\bh_i,\theta_e)), \theta_g)) - \TM_i \|_2^2
\end{equation}
using the Adam algorithm \citep{Kingma2014} with a learning rate of $1e^{-4}$, a batch size of $32$, over $1000$ epochs. We split our dataset into $85\%$ training and $15\%$ validation for $10,000$ models. A visual depiction of the network architecture is shown in Figure~\ref{fig:network} with further descriptions in Table~\ref{tab:network_table}.

\begin{table} [!htb]
    \centering
    \begin{tabular}{ccc p{0.3\linewidth}}
       Network  & Input & Output & Description\\
       \hline
       $E (\bh,\theta_e)$  & $n_s$ & $n_c \times n_t \times n_s$ & Flight height embedding\\ \hline
       $F ((\TD, E (\bh,\theta_e)), \theta_f)$  & $n_c \times n_t \times n_s $ & $n_c \times n_z \times n_s$  & Time-to-depth conversion\\ \hline
       $G(F ((\TD, E (\bh,\theta_e)), \theta_g))$  & $n_c \times n_z \times n_s$ & $n_z \times n_s$ & ResNet to process intermediate features into conductivities \\ \hline
    \end{tabular}
    \caption{Description of the networks.}
    \label{tab:network_table}
\end{table}

\section{Training Models and Data}

To generalize a network, the training data from a collection of synthetic models with known conductivities must have similar attributes to those found in the field dataset, where the true conductivities are unknown. The end-goal of the training is the ability to use the network to invert a large AEM field dataset within seconds. Our specific interest is aquifer-bearing regions, and a Skytem \citep{Sorensen2004} AEM dataset from the Kaweah sub-basin in California serves as our test field data, and prior inversion results from this dataset can be found in \cite{Kang2022}.

\begin{figure}[!htb]
	\centering
	\includegraphics[width=0.8\textwidth]{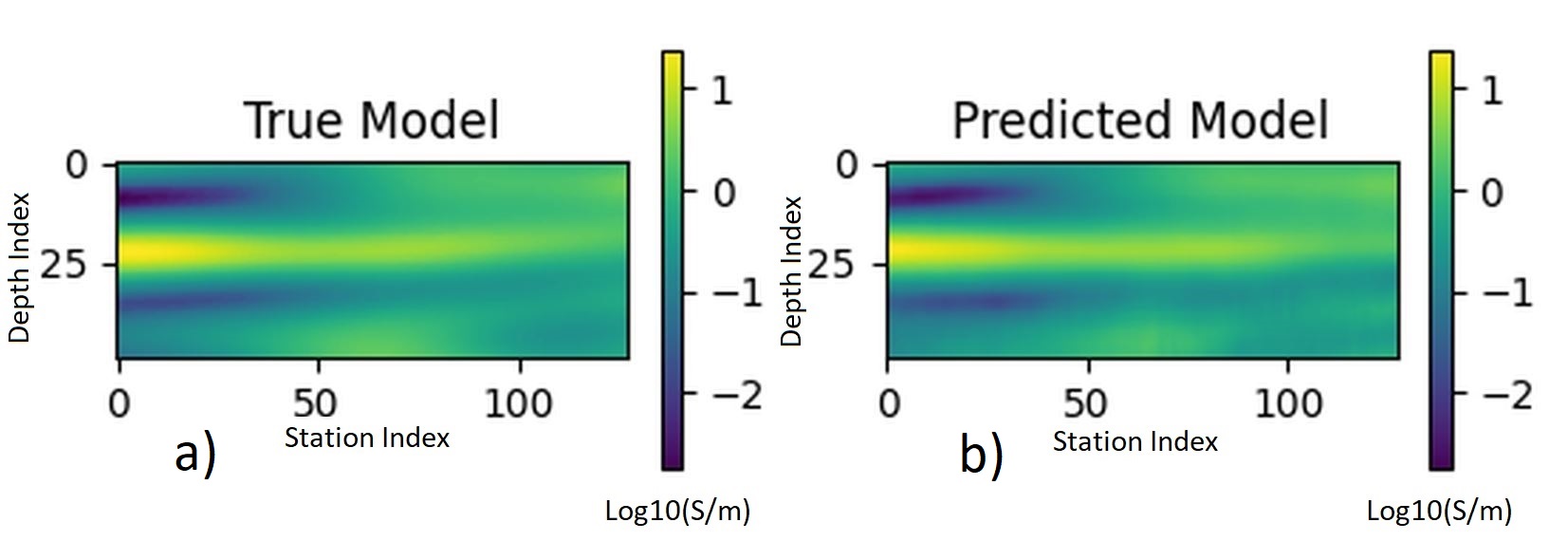}
	\caption{Validation example after training. a) True model. b) Predicted model from validation.}
	\label{fig:validation8}
\end{figure}

Our training model set consists of $10,000$ conductivity models with a core size of fine cells of 1280m (width) x 245m (deep), with a mesh size of 10m x 10m x 5m, and an example is shown in Figure~\ref{fig:validation8}a. Using the Python gstools library \citep{Muller2022} we generate random models with structures that are plausible for aquifer bearing geology. To obtain appropriate conductivity distributions for the training models, we first invert the field data using a conventional 1D inversion algorithm (EM1DTM) \citep{Farquharson1996}. Then we use the same conductivity statistics from the 1D inversions to create the training models. While the limitations of 1D inversions are well known, our thesis is that the global distribution of conductivity values is sufficiently accurate to use in building the training models. 

The training data are then calculated by forward modeling in 3D high and low-moment Skytem data with H3DTD \citep{Haber2007}. The AEM data, flight heights and labels are all normalized by the standard deviation prior to entering the neural network, while the labels also have the mean subtracted before normalization. The network is first trained and then predictions are made on the high and low-moment field data. 

\section{Results}

\begin{figure}[!htb]
	\centering
	\includegraphics[width=\textwidth]{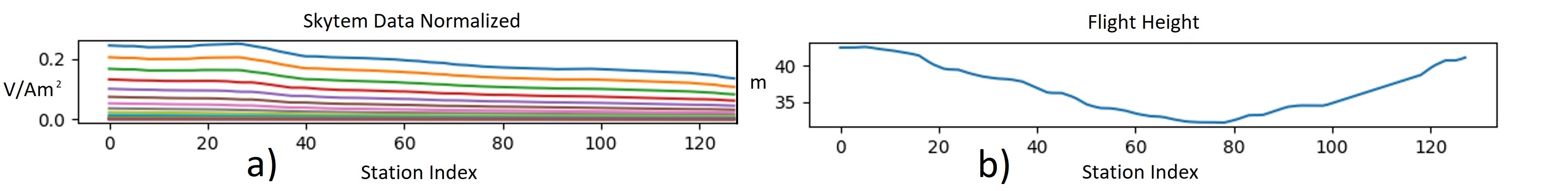}
	\caption{Field data example from Kaweah. Top) High-moment normalized data. Bottom) Flight heights.}
	\label{fig:caliData}
\end{figure}

\begin{figure}[!htb]
	\centering
	\includegraphics[width=\textwidth]{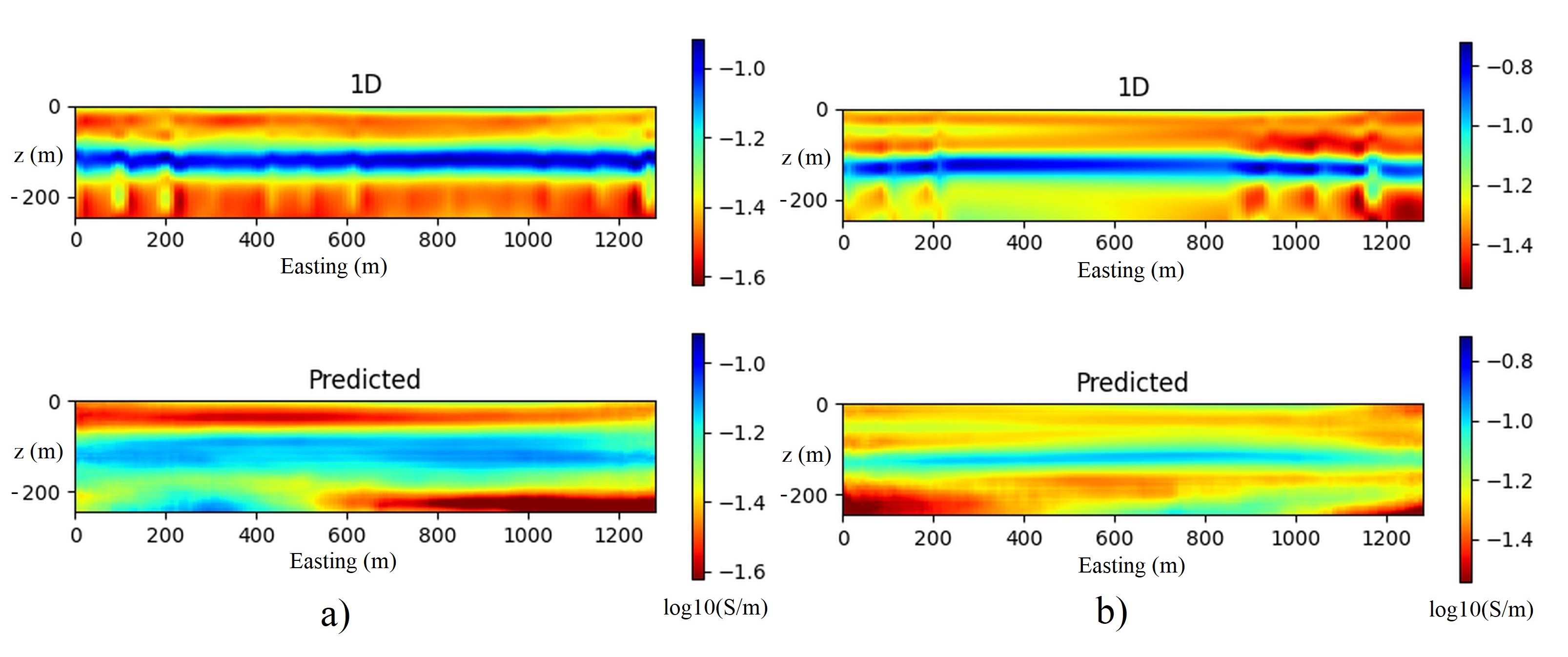}
	\caption{Network predictions from two line sections from the Kaweah sub-basin dataset. Results from stitched 1D inversions are shown in the top panel and from the trained network in the bottom panel. Note: z = 0m represents the top of the model and not sea-level.}
	\label{fig:prediction}
\end{figure}

In training, the validation loss flattened out at roughly 800 epochs to a loss of 0.1, and an example of a typical validation result is shown in Figure~\ref{fig:validation8}b, where the validation is close to the true image. We then tested the network on the Kaweah field dataset where an example of the normalized input data (high-moment) along with the raw flight heights from a line section is shown in Figure~\ref{fig:caliData}. Network predictions, from two line sections along with stitched-1D inversions for reference are displayed in Figure~\ref{fig:prediction}. 

\begin{figure}[!htb]
	\centering
	\includegraphics[width=\textwidth]{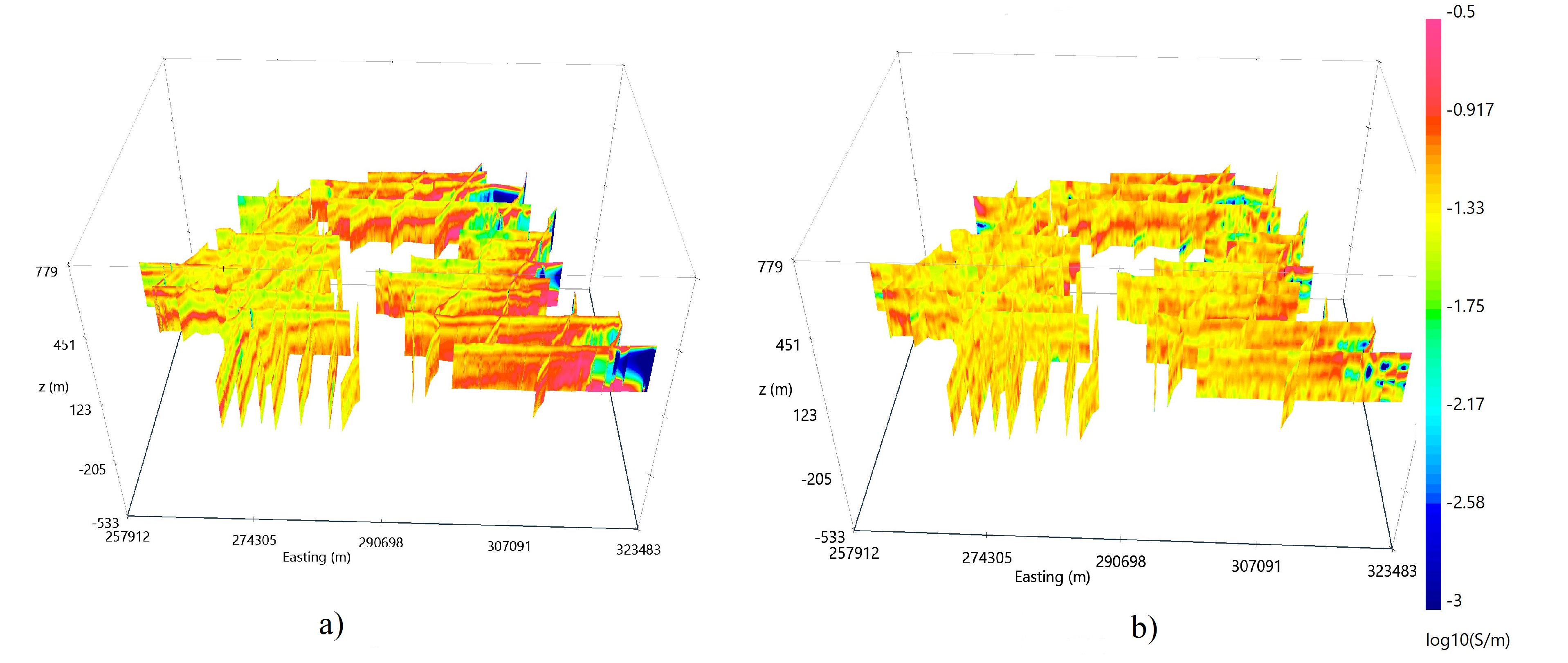}
	\caption{Skytem Kaweah dataset results a) Stitched 1D inversions. b) Neural network predictions.}
	\label{fig:predictions_All}
\end{figure}

It is clear that the network predictions have many similarities to the 1D inversions, which suggests that the predictions are producing reasonable results. The exact conductivities are slightly different as are the precise geometries of the horizontal layers, but the overall look and feel is similar. For a broader look at all the results, the stitched 1D inversions and network predictions are displayed in Figure~\ref{fig:predictions_All}. The predictions, which take only a matter of seconds to create, produce similar conductivity structures in each line compared to the 1D results, but the stitched 1D inversions produce horizontal layers that are better defined and with more resolution. Therefore, to achieve the same resolution, and eventually even better, compared to the stitched 1D results, we need more training models. $10,000$ is still a tiny number of training models in the machine learning space, and we expect the prediction results to improve as we increase this number, with the added benefit of avoiding 1D pant-leg artifacts with vertical conductors.

\section{Conclusions}

In this study we trained a neural-network to predict 2D conductivity models from time-domain AEM data, thus bypassing the need for expensive forward modeling during inversion. To train the network, $10,000$ synthetic models were created with similar conductivity attributes to the field data environment and synthetic AEM data were calculated in 3D over these training models. These data were used to train and validate the network, and then we predicted on field data from Kaweah in California and compared the predictions to stitched 1D inversion results. The AI predictions compared well with the 1D results, but did not quite show as much resolution in the horizontal layering. We see this as a successful proof of concept and we are confident that the results will improve with the addition of more training models.

\normalsize
\bibliography{bibliography}

\begin{thebibliography}{12}
\providecommand{\natexlab}[1]{#1}
\providecommand{\url}[1]{\texttt{#1}}
\expandafter\ifx\csname urlstyle\endcsname\relax
  \providecommand{\doi}[1]{doi: #1}\else
  \providecommand{\doi}{doi: \begingroup \urlstyle{rm}\Url}\fi

\bibitem[Farquharson and Oldenburg(1996)]{Farquharson1996}
C.~G. Farquharson and D.~W. Oldenburg.
\newblock Approximate sensitivities for the electromagnetic inverse problem.
\newblock \emph{Geophysical Journal International}, 126:\penalty0 235--252, 7
  1996.
\newblock ISSN 0956540X.
\newblock \doi{10.1111/j.1365-246X.1996.tb05282.x}.
\newblock URL
  \url{http://gji.oxfordjournals.org/cgi/doi/10.1111/j.1365-246X.1996.tb05282.x}.

\bibitem[Wilson et~al.(2006)Wilson, Raiche, and Sugeng]{Wilson2006}
G.~A. Wilson, A.~P. Raiche, and F.~Sugeng.
\newblock 2.5d inversion of airborne electromagnetic data.
\newblock \emph{Exploration Geophysics}, 37:\penalty0 363--371, 2006.
\newblock ISSN 18347533.
\newblock \doi{10.1071/EG06363}.

\bibitem[Haber et~al.(2007)Haber, Oldenburg, and Shekhtman]{Haber2007}
Eldad Haber, Douglas~W. Oldenburg, and R.~Shekhtman.
\newblock Inversion of time domain three-dimensional electromagnetic data.
\newblock \emph{Geophysical Journal International}, 171:\penalty0 550--564, 11
  2007.
\newblock ISSN 0956540X.
\newblock \doi{10.1111/j.1365-246X.2007.03365.x}.
\newblock URL \url{http://doi.wiley.com/10.1111/j.1365-246X.2007.03365.x}.

\bibitem[McMillan et~al.(2015)McMillan, Schwarzbach, Haber, and
  Oldenburg]{McMillan2015}
Michael~S. McMillan, Christoph Schwarzbach, Eldad Haber, and Douglas~W.
  Oldenburg.
\newblock 3d parametric hybrid inversion of time-domain airborne
  electromagnetic data.
\newblock \emph{Geophysics}, 80:\penalty0 K25--K36, 9 2015.
\newblock ISSN 0016-8033.
\newblock \doi{10.1190/geo2015-0141.1}.
\newblock URL \url{http://library.seg.org/doi/10.1190/geo2015-0141.1}.

\bibitem[Wu et~al.(2022)Wu, Huang, and Zhao]{Wu2022}
Sihong Wu, Qinghua Huang, and Li~Zhao.
\newblock Instantaneous inversion of airborne electromagnetic data based on
  deep learning.
\newblock \emph{Geophysical Research Letters}, 49, 5 2022.
\newblock ISSN 19448007.
\newblock \doi{10.1029/2021GL097165}.

\bibitem[Ronneberger et~al.(2015)Ronneberger, Fischer, and
  Brox]{Ronneberger2015}
Olaf Ronneberger, Philipp Fischer, and Thomas Brox.
\newblock U-net: Convolutional networks for biomedical image segmentation.
\newblock \emph{International Conference on Medical Image Computing and
  Computer-Assisted Intervention}, pages 234--241, 5 2015.
\newblock URL \url{http://arxiv.org/abs/1505.04597}.

\bibitem[He et~al.(2016)He, Zhang, Ren, and Sun]{He2016}
Kaiming He, Xiangyu Zhang, Shaoqing Ren, and Jian Sun.
\newblock Deep residual learning for image recognition.
\newblock \emph{Proceedings of the IEEE conference on computer vision and
  pattern recognition}, pages 770--778, 2016.
\newblock URL \url{http://image-net.org/challenges/LSVRC/2015/}.

\bibitem[McMillan et~al.(2021)McMillan, Haber, Peters, and
  Fohring]{McMillan2021}
Michael~S. McMillan, Eldad Haber, Bas Peters, and Jennifer Fohring.
\newblock Mineral prospectivity mapping using a vnet convolutional neural
  network.
\newblock \emph{Leading Edge}, 40:\penalty0 99--105, 2 2021.
\newblock ISSN 19383789.
\newblock \doi{10.1190/tle40020099.1}.

\bibitem[Kingma and Ba(2014)]{Kingma2014}
Diederik~P. Kingma and Jimmy Ba.
\newblock Adam: A method for stochastic optimization.
\newblock \emph{arXiv preprint arXiv:1412.6980}, 12 2014.
\newblock URL \url{http://arxiv.org/abs/1412.6980}.

\bibitem[Sørensen and Auken(2004)]{Sorensen2004}
Kurt~I Sørensen and Esben Auken.
\newblock Skytem – a new high-resolution transient electromagnetic system.
\newblock \emph{Exploration Geophysics}, 35:\penalty0 191--199, 2004.
\newblock ISSN 0812-3985.
\newblock \doi{10.1071/EG04194}.

\bibitem[Kang et~al.(2022)Kang, Knight, and Goebel]{Kang2022}
Seogi Kang, Rosemary Knight, and Meredith Goebel.
\newblock Improved imaging of the large-scale structure of a groundwater system
  with airborne electromagnetic data.
\newblock \emph{Water Resources Research}, 58, 4 2022.
\newblock ISSN 19447973.
\newblock \doi{10.1029/2021WR031439}.

\bibitem[Müller et~al.(2022)Müller, Schüler, Zech, and Heße]{Muller2022}
Sebastian Müller, Lennart Schüler, Alraune Zech, and Falk Heße.
\newblock Gstools v1.3: a toolbox for geostatistical modelling in python.
\newblock \emph{Geoscientific Model Development}, 15:\penalty0 3161--3182,
  2022.

\end{thebibliography}

\clearpage

\end{document}